\documentclass[journal=jctcce,manuscript=article,layout=traditional]{achemso}
\usepackage{helvet}
\usepackage{booktabs}

\newcommand{\shole}{$\sigma$-hole}

\title{The Effect of Off-Center $\sigma$-Hole on the Atom-Centered Partial Charges in Halogenated Molecules}

\author{Aneta Leskourov\'a}
\affiliation{Department of Physical Chemistry, University of Chemistry and Technology, Technická 5, 16628 Prague, Czech Republic}

\author{Michal H. Kol\'a{\v r}}
\email{michal@mhko.science}
\affiliation{Department of Physical Chemistry, University of Chemistry and Technology, Technická 5, 16628 Prague, Czech Republic}

\begin{document}

\singlespacing

\begin{abstract}
Partial atomic charges belong to key concepts of computational chemistry. In some cases, however, they fail in describing the electrostatics of molecules. One such example is the \shole, a region of positive electrostatic potential located on halogens and other atoms. In molecular mechanics, the \shole ~is often modeled as a pseudo-atom with a positive partial charge located off the halogen nucleus. Here we address a question, to what extent the pseudo-atom affects partial charges of other atoms in the molecule. To this aim, we have thoroughly analyzed partial charges of over 2300 halogenated molecules from the ZINC database calculated by the Restricted Electrostatic Potential (RESP) method and compared them with the charges fitted by RESP including the pseudo-atom. We show that the pseudo-atom improves charge fitting for a vast majority of molecules. The \shole, modeled as the off-center charge, affects the atoms within three covalent bonds from the halogen.
\end{abstract}

\maketitle

\section{Introduction}

Chemists have dealt with partial atomic charges for centuries. The notion of a positive or negative character of an atom was formalized by the introduction of electronegativity in the early 19th century \cite{Jensen96}. Although there is no physical observable corresponding to the partial atomic charges, they have become one of the key concepts in chemistry. Yet, on some occasions, the concept fails.

For instance, the description of the carbonyl oxygen electrostatics by a single partial charge is inaccurate due to electron lone pairs on the oxygen. A systematic study on the deficiency of atom-centered partial charges was done by Kramer et al. \cite{Kramer14}. On a set of 65 representative organic compounds, they showed that atom-centered multipoles up to the quadrupole improve electrostatic potential (ESP) when compared to ESP values obtained from electric monopoles, i.e. the partial charges. Compared to the reference quantum mechanical (QM) ESP, the multipoles were found essential especially for nitrogen, sulfur and halogen atoms.

More generally, partial charges fail when the ESP around an atom is anisotropic. Apart from lone pairs, so-called \shole s have attracted attention in the recent decade or two. The \shole ~appears on halogen atoms \cite{Brinck92,Clark07,Mallada21}, but it was also identified on chalcogens \cite{Vogel19}, pnictogens \cite{Bauza13}, or even tetrels \cite{Bauza16}. The \shole ~represents an elegant explanation of attractive noncovalent interactions known as halogen bonds \cite{Desiraju13, Cavallo16}, or more broadly \shole ~bonds \cite{Lim18}. These interactions have found applications in crystal design \cite{Metrangolo08}, catalysis \cite{Bulfield16}, drug development \cite{Mendez17}, supramolecular chemistry \cite{Gilday15} and more. A large amount of information about \shole s has also been collected through computational techniques. \cite{Kolar16}.

Computational and theoretical chemists rely on partial charges in molecular mechanical (MM) studies, classical molecular dynamics simulations, free-energy calculations, molecular docking, etc. Typically, the partial atomic charges of a molecule are fitted to reproduce reference quantum chemical data, or experimental quantities such as density or relative permittivity \cite{Wiberg93}. A particularly successful implementation of the fitting procedure is the Restricted Electrostatic Potential (RESP) method \cite{Bayly93} which restrains the magnitude of the charges to 0.0 $e$. A charge model was later derived based on the QM method AM1 enhanced by so-called bond charge correction (BCC) \cite{Jakalian00, Jakalian02}. The AM1-BCC charges have gained popularity due to low computational demands and accuracy comparable to the RESP charges.

For a more realistic description of anisotropic ESP within non-polarizable MM potentials, off-atom partial charges have been introduced through pseudo-atoms. A prominent example is the four-center TIP4P water model \cite{Jorgensen83}, where an extra site, located inside the triangle of oxygen and two hydrogen atoms, improves the properties of liquid water compared to the three-center TIP3P model.

The off-atom charges were introduced to improve hydrogen bonding of sulfur in the early AMBER force field \cite{Weiner84}. So far, using off-center charges has not reached the popularity of the atom-centered force fields, however. Dixon and Kollman added negative pseudo-atoms to simple organic compounds to mimic the electron lone pairs of N, O, and S. This improved ESP at MM level, and angular dependence of hydrogen bonds. Later, the pseudo-atoms were used to enhance sulfur atoms in the OPLS-AA force field \cite{Yan17}.

ESP anisotropies are usually well localized in space. For instance, the \shole ~position on the halogen is not affected by the chemical environment. An analysis of about 2500 molecules from the ZINC database of organic drug-like compounds \cite{Irwin12} revealed that the angular deviation of the \shole s from the straight direction defined by the C--X bond is typically smaller than 5$^\circ$ \cite{Kolar14}.

For small molecules like bromomethane, we would expect that the charge of the pseudo-atom introduced to the MM model to mimic the \shole ~affects all other partial charges. For larger molecules like drugs, fluorescent dyes, or biomolecular building blocks, it remains unclear which atom-centered partial charges would be affected by introducing the pseudo-atom and how. This question could be formulated in terms of the inductive effect widely used by organic chemists to explain mechanisms of chemical reactions \cite{Ingold34, Wheland35}. Due to spatial limits of the inductive effect, we would expect only the nearest atomic charges affected by the \shole ~charge.

\begin{figure}
    \centering
    \includegraphics{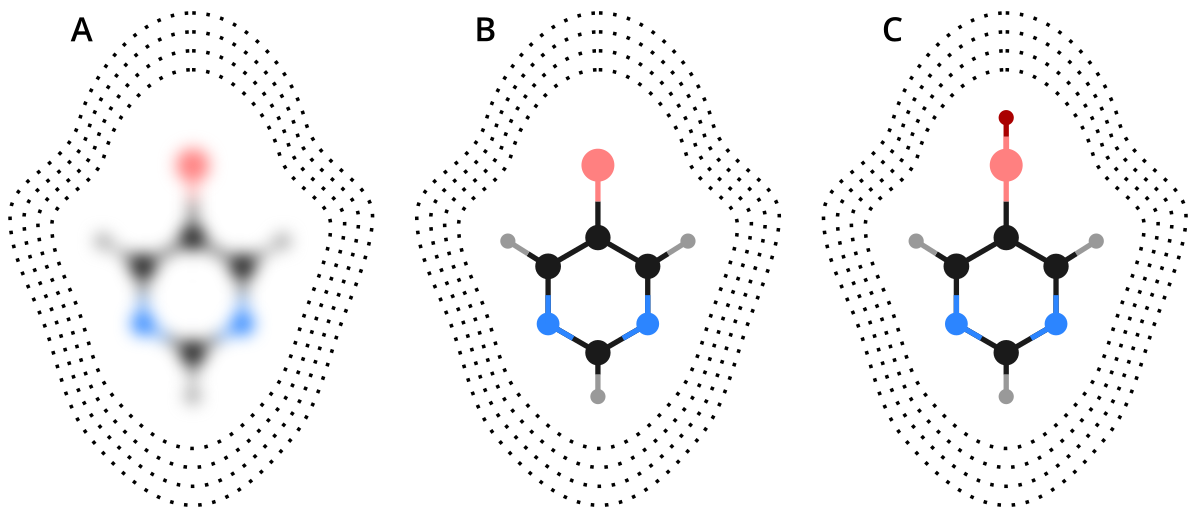}
    \caption{A scheme of charge fitting for 5-iodopyrimidine (ZINC00967270).  Atoms are color coded as follows: C black, N blue, I pink, H gray, pseudo-atom deep red. Dotted lines represent the ESP grid. A) The quantum chemical nature of atoms is represented by blurred circles. B) RESP with atom centered point charges. C) mRESP with the pseudo-atom on iodine.}
    \label{fig:scheme}
\end{figure}

Here we address the question about the effect of the pseudo-atom quantitatively. By comparing charge distributions obtained by the RESP method for molecules including and excluding the pseudo-atom on halogen (Fig.~\ref{fig:scheme}), we assess the \shole ~counterpart of the inductive effect. We have done it for a set of more than 2300 drug-like monohalogenated molecules to guide the reader's intuition about changes brought by the off-atom partial charge.

\section{Methods}

\subsection{Dataset}

Molecules used for this study were obtained from the database of drug-like molecules for virtual screening ZINC \cite{Irwin12}. We selected commercially available neutral molecules with exactly one atom of chlorine, bromine, or iodine. In the database query, we limited the molar mass to 250 Da, 300 Da, and 350 Da, for chlorinated, brominated, and iodinated molecules, respectively. The limit was chosen to avoid a strong dependence of charges on conformations, as could be expected due to the many conformational states of large molecules. In all molecules, the halogen was covalently bound to a carbon atom. Altogether, the dataset contains 2311 molecules. A similar dataset was used previously to study \shole ~properties \cite{Kolar14}. The summary of the molecules investigated is in Table~\ref{tab:dataset}. The database accession codes of the molecules are provided in the Supplementary Information.

\begin{table}[]
    \centering
    \begin{tabular}{lcccc}
    \toprule
     & Cl & Br & I & whole set  \\
    \midrule
    number of molecules                & 276 & 1120 & 915 & 2311 \\
    fraction of total set / \%         & 12  & 49   & 39  & 100 \\
    median of number of atoms          & 26  & 24   & 27  & 26 \\
    fraction$^a$ of sp$^3$ C$_1$ / \%  & 10  & 16   & 2   & 10 \\
    \bottomrule
    \multicolumn{5}{l}{\footnotesize $^a$the remaining C$_1$ atoms were sp$^2$ hybridized} \\
    \end{tabular}
    \caption{Dataset overview}
    \label{tab:dataset}
\end{table}

\subsection{QM Calculations}

All molecules were energy minimized as described previously \cite{Kolar14} using the Gaussian 09 program package \cite{Frisch09}. In brief, B3LYP \cite{Becke93,Lee88} with 6-31G* basis set \cite{Hehre72,Hariharan73} was used for all atoms. The exception was iodine, for which the LANL2DZ basis set was used together with the pseudopotentials for the inner-core electrons to account for scalar relativistic effects \cite{Wadt85}. For each molecule, all subsequent charge fitting was performed on the energy minimized geometry. The ESP was calculated using Hartree-Fock with the same basis set as for the energy minimization. This workflow and the level of theory were originally used for the charge derivation by RESP \cite{Bayly93} and it is compatible with the several force fields from the AMBER family. The level of theory is only approximate, which is however not a problem here, where we use it as a reference of mutual comparison of two MM methods.

The ESP grid was constructed using 5 layers with the density of 2 points per square bohr (IOPs 6/41 and 6/42, respectively) resulting on average in about 3300, 3700, 4400 grid points for chlorinated, brominated, and iodinated molecules, respectively. 

\subsection{Charge Fitting}

The partial atomic charges were computed by the RESP method \cite{Bayly93} as implemented in the \emph{antechamber} program from the Amber 14 suite \cite{Case14}. The antechamber reads the Gaussian output file and yields a set of files containing various information about the fitting process. As in the default version, RESP was a two-stage optimization of partial charges at the atomic positions. In the first stage, all charges were optimized without constraints. The initial guess was 0.0 \emph{e} for all atoms. In the second stage, non-polar atoms were re-optimized with the constraints accounting for the chemical symmetry \cite{Bayly93}. 

To mimic the \shole, we added a pseudo-atom to the RESP procedure, hereafter denoted modified RESP (mRESP). In practice, this was done by modifying the output files of the standard run of the antechamber and re-running the RESP by the \emph{resp} module of the antechamber. The pseudo-atom fitting center was added in the elongation of the C--X bond to the distance of 1.225, 1.295, 1.386 \AA ~from the chlorine, bromine, and iodine, respectively (0.7 times the halogen van der Waals radius). It was argued previously \cite{Kolar12} that the pseudo-atom position and its partial charge correlate. Thus it is rather arbitrary, where to place the pseudo-atom given it stays within the van der Waals sphere.

The two-stage nature of mRESP procedure was kept identical to the standard RESP. The pseudo-atom charge was set to 0.0 \emph{e} in the beginning and was free to change in both fitting stages.

\subsection{Analyses}

We compared the partial charges obtained by fitting with and without the off-center pseudo-atom. The difference of the partial charges $\Delta q$ of an atom A was defined

\begin{equation}
    \Delta q(\mathrm{A}) = q_m(\mathrm{A}) - q(\mathrm{A}),
\end{equation}
where the subscript \emph{m} denotes the value obtained by the mRESP.

Further, we calculated dipole moments $\mu$ and compared them with the QM values at the HF/6-31g* level. Although the dipole moments at this level of theory are only approximate, they still allowed us to assess how the electrostatics of the molecules, given by the reference ESP grid, is reproduced at the MM level. The quality of the ESP fit was compared by the sum of square deviations of the ESP values at MM level calculated by RESP or mRESP ($V_{\mathrm{MM}}$) and the reference QM values obtained by Gaussian ($V_{\mathrm{QM}}$)

\begin{equation}
    \chi^2 = \sum_i^N (V_{\mathrm{MM}} - V_{\mathrm{QM}})^2,
\end{equation}
where the sum runs over N grid points for which the ESP was calculated. The $\chi^2$ and $\chi_m^2$ were obtained directly from the antechamber output files as the "residual sum of squares" for RESP and mRESP runs, respectively.

To evaluate the role of the chemical environment of the halogen on the \shole, we assigned the atom types defined by the General Amber Force Field (GAFF) \cite{Wang04}. The assignment was done during the standard antechamber run (module \emph{atomtype}). We numbered the atoms according to their topological distance from the halogen. The nearest carbon atom C$_1$ within one covalent bond from the halogen was classified as sp$^2$ or sp$^3$ hybridized based on GAFF atom types. Some of the analyses were done for the subsets of the molecules according to the C$_1$ hybridization.

\section{Results and Discussion}

%\subsection{RESP with the off-center \shole ~outperforms standard RESP}

First, we quantify the quality of the fit by the mRESP method, i.e. including the pseudo-atom mimicking the \shole. The pseudo-atom improved the fit in more than 99\% of cases compared to the standard RESP fit, as indicated by the $\chi^2$. Only for 6 brominated and 4 iodinated molecules, the presence of the pseudo-atom during the ESP fit led to a deterioration of $\chi^2$. In eight of those cases, the halogen was bound to an sp$^3$ hybridized carbon atom.

The $\chi^2$ being lower for mRESP than standard RESP goes in line with the notion that a better fit is obtained when more fitting parameters are used. Here, the use of pseudo-atom increases the number of fitted parameters by one. Generally, increasing the number of parameters may lead to an overfitted model. Overfitting is, however, not an issue with RESP, where the size of the reference data -- ESP grid of QM values -- is often two orders of magnitude larger than the number of parameters fitted. 

As a second quality measure of the fit, the molecular dipole moments were compared. The effect of the pseudo-atom on the dipole moments was quantified by the absolute deviation of $\mu$ from $\mu_\mathrm{QM}$. The mRESP yielded dipole moments closer to the reference QM values than standard RESP in more than 78\% of the molecules. For heavier halogens, the improvement was more frequent than for the lighter ones, namely 70, 79, and 81\% for chlorinated, brominated, and iodinated molecules.

%\subsection{Halogen with the pseudo-atom is more negative on average}

\begin{figure}
    \centering
    \includegraphics{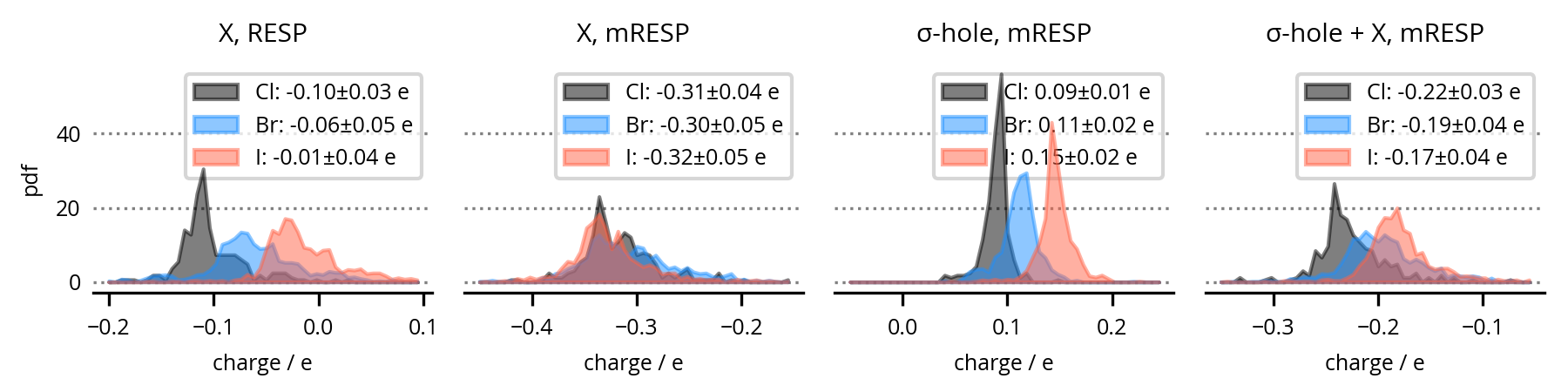}
    \caption{Probability density functions (pdf) of selected partial charges obtained by standard RESP or modified RESP (mRESP). In the legend, the mean values and standard deviations of the pdfs are given. The x-axes have the range of 0.3\,$e$ in all cases.}
    \label{fig:charges}
\end{figure}

Next, we analyze selected partial charges. Fig.~\ref{fig:charges} shows normalized histograms of partial charges for the halogen and \shole ~for each class of halogenated molecules calculated by RESP or mRESP. The standard RESP yields halogens mostly negative. On average, the RESP partial charge is less negative for heavier halogens, following the trend of their electronegativities. However, a non-negligible fraction of molecules is modelled by standard RESP as having a positive halogen. The fraction increases with the atomic number of the halogen from about 1\%, through 11\% to 32\% for chlorine, bromine, and iodine, respectively. It suggests that on these molecules the \shole, represented by the positive ESP on the QM grid near the halogen, is positive enough to cause a positive atom-centered charge. Halogens up to iodine are more electronegative than carbon and therefore they should be negative. The observed fraction of positive halogen partial charges is a contradiction. On the other hand, it has been established that the \shole ~magnitude increases with the increasing atomic number of the halogens \cite{Clark07, Kolar14}, which agrees with the largest fraction of positive $q(X)$ for iodine. In fact, halogens exhibit ambivalent nature. They have \shole, as well as a ring-like region of negative ESP. The positive or negative atom-centered partial charge cannot accurately describe halogen electrostatic properties.

The use of pseudo-atom in charge fitting smears the differences between the halogens out. The histograms of $q_m(\mathrm{X})$ for chlorine, bromine, and iodine almost overlap with the mean values around --0.3\,$e$. Compared to standard RESP, mRESP yields systematically more negative partial charges centered on the halogen nucleus.

On the other hand, the pseudo-atom's charge fitted by mRESP is always positive, as expected for the \shole. Here, the pseudo-atom charge is on average more positive for iodine (0.15\,$e$) than for bromine (0.11\,$e$) and chlorine (0.09\,$e$). The variance in pseudo-atom charge is about one-half of the variance of the halogen charge across the studied set of molecules. It indicates that the partial charge of the pseudo-atom is well determined by the fitting procedure likely because it is located close to the reference grid of ESP points.

In the mRESP model, the sum of the atom-centered charge $q_m(\mathrm{X})$ and the off-center charge $q_m(\sigma)$ characterizes the ESP of the halogen atom. It is thus meaningful to compare it with the $q(\mathrm{X})$ of the RESP model (where only a single atom-centered charge plays this role). The histograms in Fig.~\ref{fig:charges} reveal that mRESP yields overall more negative halogen atoms than standard RESP. For all molecules, the sum of $q_m(\sigma)$ and $q_m{\mathrm{X}}$ is negative, which follows the chemical intuition guided by the electronegativities.

%\subsection{Effect of pseudo-atom vanishes with distance}

\begin{figure}
    \centering
    \includegraphics{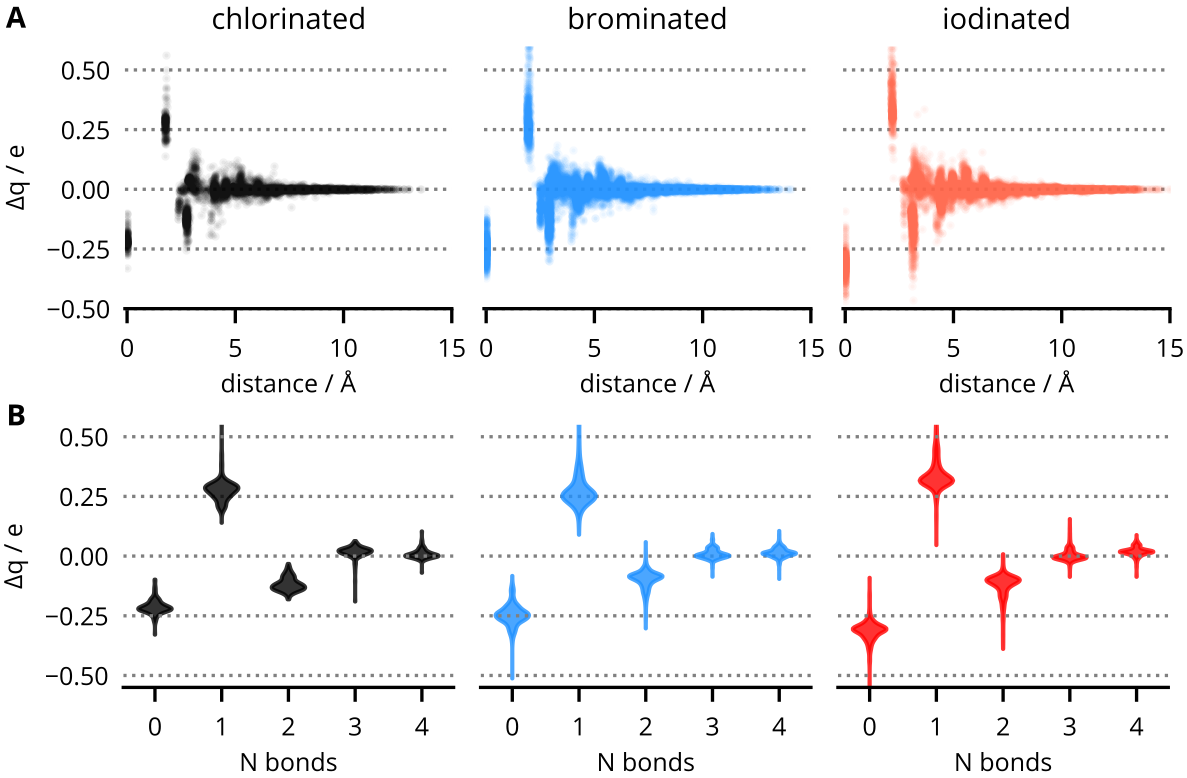}
    \caption{Difference between mRESP and RESP charges $\Delta q$ as a function of distance from the halogen atom. A) $\Delta q$ is plotted against Euclidean distance between an atom and the halogen. Each dot represents one atom. B) Violin plots of $\Delta q$ as a function of the number $N$ of bonds between an atom and the halogen.}
    \label{fig:dist}
\end{figure}

To understand, how localized the effect of pseudo-atom is, we analyzed the partial charges considering their distance from the halogen (Fig.~\ref{fig:dist}). The difference $\Delta q$ between mRESP and standard RESP is plotted as a function of the Euclidean distance from the halogen atom (Fig.~\ref{fig:dist}A). $\Delta q$ is large near the halogen atom and vanishes for larger distance. In the distance of about 7~\AA, $\Delta q$ is close to zero. This trend is alike irrespective of the halogen.

Fig.~\ref{fig:dist}B shows $\Delta q$ as a function of number of covalent bonds separating an atom and the halogen. The negative $\Delta q$ values alternate with the positive values similarly to the alternations with Euclidean distance (Fig.~\ref{fig:dist}A). The atoms within 3--4 covalent bonds from the halogen have the partial charges almost the same for the mRESP and standard RESP fitting procedures. Hence, the effect of the positive pseudo-atom is limited to the halogen, nearest carbon atom C$_1$, and the atoms bound to the C$_1$. For most molecules due to sp$^2$ hybridization of C$_1$, there are two more atoms in the distance of two covalent bonds. These results show that the effect of the pseudo-atom is rather localized.

\section{Conclusions}

In this work, we have analysed partial atomic charges of a large set of halogenated molecules. We have investigated, how the use of an off-center fitting position mimicking the \shole ~affects atom-centered partial charges. Our main finding is that the effect vanishes with distance from the halogen atom. Namely, the distance of about 7 \AA ~or 3 covalent bonds from the halogen is sufficient to consider the pseudo-atom effect negligible.

Partial charges provide a proxy to the inductive effect of the \shole. Our results show that the effect of the \shole ~is spatially limited to a few nearest atoms. Since partial charges are widely used in many computational techniques, we provide support for more accurate halogen potentials to be localized in space. 

For instance in computer-aided drug design, molecular docking of halogenated compounds was employed. While some approaches used an empirical distance-based potential function to describe halogen bonding \cite{Koebel16}, pure electrostatics with pseudo-atom was also used. Unlike more accurate MM studies, no charge fitting was performed and the off-center pseudo-atom was assigned with a fixed partial charge \cite{Kolar13} the magnitude of which was subtracted from the atom-centered partial charge of the halogen. Our findings provide support for the limited scope of the charge modifications. On the other hand, an improvement could be obtained by making the halogen with the \shole ~more negative overall and by a modification of the atom next to the halogen.

Further in molecular docking, a technique is used that divides a candidate molecule into fragments \cite{Chen09, Huang10}, which are then docked separately. Our results show that electrostatics-driven docking of a halogenated fragment with the pseudo-atom would be reasonable and the effect of the pseudo-atom to the other molecular fragments can be neglected. Only a portion of \emph{a priori} known partial charges in molecular docking \cite{Tsai08} would be necessary to modify to account for the \shole. Overall, our findings may help designing more accurate intermolecular potentials, especially suited for the modeling condensed phase.

\section{Acknowledgement}

We thank Hugo McGrath for helpful comments on the manuscript. This work was partially supported by the Czech Science Foundation through project no. \emph{19-06479Y}. The accession ZINC codes for molecules investigated, files generated during the RESP fitting for exemplary molecules, and analysis scripts are available on\\ https://github.com/mhkoscience/leskourova-offcenter.

\section{Authors Contribution}

MHK designed and supervised the research; AL and MHK performed the calculations and analyzed the results; MHK wrote the initial version of the manuscript; AL and MHK finalized the manuscript.

%\bibliography{refs-abbr.bib}
\providecommand{\latin}[1]{#1}
\makeatletter
\providecommand{\doi}
  {\begingroup\let\do\@makeother\dospecials
  \catcode`\{=1 \catcode`\}=2 \doi@aux}
\providecommand{\doi@aux}[1]{\endgroup\texttt{#1}}
\makeatother
\providecommand*\mcitethebibliography{\thebibliography}
\csname @ifundefined\endcsname{endmcitethebibliography}
  {\let\endmcitethebibliography\endthebibliography}{}

\end{document}